# PIII PROJECT OVERVIEW AND STATUS*

R. Stanek†, C. Boffo, S. Chandrasekaran, S. Dixon, E. Harms, L. Kokoska, I. Kourbanis, J. Leibfritz, O. Napoly, D. Passarelli, E. Pozdeyev, and A. Rowe Fermilab, Batavia, Illinois USA

*Abstract*

The Proton Improvement Plan II (PIP-II) Project at Fermilab is constructing an 800 MeV H- ion superconducting radio frequency linear accelerator (SRF Linac). Capitalizing on advances in superconducting radio frequency (SRF) technology, this new Linac will provide the initial acceleration chain to power the Fermilab accelerator complex throughout the next decade and beyond. In addition, upgrades to the existing Booster, Main Injector, and Recycler rings will enable them to operate at a 20 Hz repetition rate and will provide a 1.2 MW proton beam to drive neutrino research at the Deep Underground Neutrino Experiment (DUNE) in Lead, South Dakota.

The SRF Linac for PIP-II consists of twenty-three cryomodules (CM) with cavities of three different frequencies and five different shapes. PIP-II is the first U.S. accelerator project to be built with major international contributions, thus benefitting from their world-leading expertise and capabilities. U.S. National Laboratory Partners include Argonne National Lab, Lawrence Berkely Lab, Jefferson Lab, and SLAC National Accelerator Lab. International Partner Countries include India, France, Italy, United Kingdom, and Poland. The PIP-II Project is one year into the execution phase and is making good progress, but as is the case in many big projects currently being executed, is facing challenges from inflationary pressures, and supply chain delays. These issues are being addressed in collaboration with Fermilab management, our Partners, and the U.S. Department of Energy.

## INTRODUCTION

PIP-II will be a world-class SRF-based accelerator which will enhance performance and operational reliability of the Fermi National Accelerator Laboratory (FNAL) complex. It is the latest step to revitalization of the accelerators that power America's High Energy Physics Program. The science objective is to enable the accelerator complex to provide 1.2 MW proton beam on the LBNF target (upgradeable to multi-MW) and to provide a flexible, multi-user capability for a broader reaching physics program. The Linac output will be 800 MeV when injecting into the existing Booster. The machine will run in pulsed mode initially but will be fully continuous wave (CW) compatible. Performance Goals are tabulated in Table 1.

Table 1: PIP-II Performance Goals

| Performance Parameter | Value | Units |
|---|---|---|
| Delivered Beam Energy | 800 | MeV |
| Beam Particles | H- | |
| Beam Pulse Length | 0.54 | ms |
| Particles per Pulse | 6.7 | $10^{12}$ |
| Pulse repetition Rate | 20 | Hz |
| Average Beam Current | 2 | mA |
| Maximum Bunch Intensity | 1.9 | $10^8$ |
| Maximum Bunch Rep Rate | 162.5 | MHz |
| Bunch Pattern | Prog and Arbitrary | |
| RF Frequency | 162.5 harmonic | |
| Bunch Length (RMS) | < 4 | ps |
| Transverse Emittance | ≤ 0.3 | mm-rad |
| Longitudinal Emittance (RMS) | ≤ 0.3 | mm-rad |

The accelerator is being built in collaboration with several key International Partners including Bhabha Atomic Research Centre (BARC), Commissariat à l'energie atomique (CEA), Centre National de la Recherche Scientifique/Institut national de physique nucléaire et de physique des particules (CNRS/IN2P3), Istituto Nazionale di Fisica Nucleare (INFN), Inter-University Accelerator Centre (IUAC), Raja Ramanna Centre for Advanced Technology (RRCAT), Science & Technology Facilities Council UK Research and Innovation (STFC UKRI), Lodz University of Technology (TUL), Variable Energy Cyclotron Centre (VECC), Wroclaw University of Science and Technology (WUST), and Warsaw University of Technology (WUT). The collaboration is functioning very well, and Partners are committed to delivering their scope of work. [2]

Additionally, FNAL is working with other Department of Energy (DOE) labs including Lawrence Berkeley National Lab (LBNL), Argonne National Lab (ANL), Thomas Jefferson National Lab (JLab) and SLAC National Accelerator Lab on various aspects of the machine.

The scope of the project includes extensive new civil construction, the SRF-based Linac, a 2.5 kW @2K cryoplant housed in a new standalone building, all necessary power and auxiliary equipment, and the connection into the Booster to allow injection. The Linac begins with a Warm Front End followed by multiple cryomodules including a Half Wave Resonator (HWR), nine Single Spoke Resonators (two SSR1 and seven SSR2), and thirteen 650 MHz Elliptical units (nine LB650 and four HB650). The beam then progresses through the Beam Transfer Line with various stages of collimation before injecting into the Booster. [1]

---

* Work supported by the U.S. Department of Energy, Office of Science, Office of High Energy Physics, under U.S. DOE Contract No. DE-AC02-07CH11359
† rstanek@fnal.gov   FERMILAB-CONF-23-331-PIP2

Upgrades to the Booster, Recycler Ring (RR) and Main Injector (MI) are also part of the Project. In particular, the Booster repetition rate will be increased from 15 Hz to 20 Hz. The injection time will be longer (550 microsec) and the area for injection will still be very tight, which requires new designs for magnets and the injection girder. The Main Injector will get an increase in Radiofrequency (RF) power by installing additional power tubes in the modulators in each of the twenty RF accelerating stations and updating the Solid-State drivers. The complex will accelerate 50% more beam in the Booster and Main Injector while reducing the (beam) power loss. [4]

## DESIGN STRATEGY

As part of the PIP-II Project Execution Plan, a design "Tailoring Strategy" was developed and approved by DOE. Readiness for project execution (DOE Critical Decision 3, CD-3) is defined as substantial completion of the final design of the Linac Complex civil construction, designs for Critical Technical Systems being sound and sufficiently mature to start civil construction, and designs being independently reviewed and well documented. Critical Technical Systems are defined as systems having the largest impact on the Linac Complex Conventional Facilities (CF) design. Excluding statutory building regulations, these PIP-II Critical Technical Systems completely determine the CD-3 CF scope. This allowed the construction of the Linac Complex to proceed without delays from other technical systems,

For the SRF cryomodules, a single prototype will be fabricated and tested. The schedule allows for some design iteration between prototype testing and the start of production. Other complex technical systems, such as magnets, high power RF amplifiers, and instrumentation, incorporate prototype or pre-series assemblies as well.

Designs are driven by Technical Requirement Specifications (TRS). Changes to the Technical Baseline are tracked per the Configuration Management Plan. All significant design changes must be processed through a Design Change Request (DCR) and approved by the Design Change Board, chaired by the Technical Integration Manager. Consistent with the Design Tailoring Strategy, initial deigns were constrained by a "Not-to Exceed" envelope which allowed civil facility design and award to proceed. As designs were established, reviewed, and approved, the actual CAD models were incorporated into the overall Linac CAD model to ensure compatibility.

Per the Systems Engineering Management Plan, the Design Authority is the Level 2 Subsystem Manager, while the governing institution responsible for final design of a component is the "Designer of Record". Current assignments for the Designer of Record are shown in Table 2.

Table 2: Designer of Record

| Item | Designer of Record | Production Entity |
|---|---|---|
| **Cryoplant** | | |
| Cryoplant | Commercial | DAE - BARC |
| **CDS** | | |
| Distribution Valve Box | WUST | WUST |
| Tunnel Transfer Line Modules | WUST | WUST |
| **SSR1 CM1-2** | | |
| CM | FNAL | FNAL |
| Jacketed Cavities | FNAL | FNAL, DAE - BARC |
| Couplers | FNAL | FNAL, DAE - BARC |
| Tuners | FNAL | FNAL, DAE - BARC |
| SC Solenoids | DAE - BARC | DAE - BARC |
| **SSR2 CM1-7** | | |
| CM (including shipping) | FNAL, DAE - BARC | FNAL, DAE - BARC |
| Jacketed Cavities | FNAL | FNAL, IN2P3, DAE - BARC |
| Couplers | FNAL | FNAL, DAE - BARC |
| Tuners | FNAL | FNAL, DAE - BARC |
| SC Solenoids | DAE - BARC | DAE - BARC |
| **LB650 CM1-9** | | |
| CM (including shipping) | CEA | FNAL & CEA |
| Jacketed Cavities | FNAL, INFN | INFN, DAE - VECC |
| Couplers | FNAL | FNAL, DAE - BARC |
| Tuners | FNAL | FNAL, DAE - VECC |
| **HB650 CM1-6** | | |
| CM (including shipping) | FNAL, CEA, DAE - RRCAT, UKRI | FNAL, UKRI, DAE - RRCAT |
| Jacketed Cavities | FNAL | UKRI, DAE - RRCAT |
| Couplers | FNAL | UKRI, DAE - BARC |
| Tuners | FNAL | UKRI, DAE - RRCAT |
| **HPRF** | | |
| 325 MHz - 7 kW SSAs (SSR1) | DAE - BARC | DAE - BARC |
| 325 MHz - 20 kW SSAs (SSR2) | DAE - BARC | DAE - BARC |
| 650 MHz - 40 kW SSAs (LB650) | DAE - RRCAT | DAE - RRCAT |
| 650 MHz - 70 kW SSAs (HB650) | DAE - RRCAT | DAE - RRCAT |
| **Magnets** | | |
| PIP-II Magnets | DAE - BARC | DAE - BARC |
| 650 MHz Warm unit magnets (quads + correctors) | DAE - BARC | DAE - BARC |

## COLLABORATION

Building a strong collaboration is an essential part of many modern-day large science projects, and PIP-II is no exception. Initial lessons learned from the PIP-II project collaboration include:

- Define a collaborative relationship upfront and develop a management and evaluation style that reflects the core principle of "we are in this together". For instance, Partners must have their voices heard and be part of the management team, otherwise it can become a customer/vendor relationship. Also, in a collaborative model there is

- a shared responsibility to deliver the scope and solve problems, taking a fully integrated team approach. Communication is the key.
- Each Party should be able to point to a benefit they are receiving from the Partnership to have strong arguments to their funding agencies for continued support.

The PIP-II Partnerships are key to the project's success and are functioning very well. This is a testament to the collaborative spirit of the lead Technical Coordinators and the focus of the Subproject Managers and Coordinators on getting the work done. The scope of each Partners work is shown in Table 3.

Table 3: Partner Scope

| Item | DAE | CEA | IN2P3 | INFN | UKRI | WUST |
|---|---|---|---|---|---|---|
| **Cryoplant** | | | | | | |
| Cryoplant | 1 | | | | | |
| **CDS** | | | | | | |
| Distribution Valve Box | | | | | | All |
| Tunnel Transfer Line Modules | | | | | | All |
| **SSR1 CM1-2** | | | | | | |
| CM | | | | | | |
| Jacketed Cavities | 9 | | | | | |
| Couplers | 9 | | | | | |
| Tuners | 9 | | | | | |
| SC Solenoids | All | | | | | |
| **SSR2 CM1-7** | | | | | | |
| CM (including shipping) | 1 kit | | | | | |
| Jacketed Cavities | 5 | | | All Tests | | |
| Couplers | 5 | | | | | |
| Tuners | 5 | | | | | |
| SC Solenoids | All | | | | | |
| **LB650 CM1-9** | | | | | | |
| CM (excluding shipping) | | 10 | | | | |
| Jacketed Cavities | 4 | | | | 40 | |
| Couplers | 4 | | | | | |
| Tuners | 4 | | | | | |
| **HB650 CM1-6** | | | | | | |
| CM (including shipping) | 1 kit | | | | 3 | |
| Jacketed Cavities | 20 | | | | 18 | |
| Couplers | 20 | | | | 18 | |
| Tuners | 20 | | | | 18 | |
| **HPRF** | | | | | | |
| 325 MHz - 7 kW SSAs (SSR1) | 9 | | | | | |
| 325 MHz - 20 kW SSAs (SSR2) | 40 | | | | | |
| 650 MHz - 40 kW SSAs (LB650) | 40 | | | | | |
| 650 MHz - 70 kW SSAs (HB650) | 31 | | | | | |
| **Magnets** | | | | | | |
| PIP-II Magnets | All | | | | | |
| 650 MHz Warm unit magnets (quads+correctors) | All | | | | | |

## SAFETY AND QUALITY CONTROL

Safety of the workers must be the first item on everyone's agenda. Ensuring that the staff goes home safe and sound, is the number one priority. Overall, the PIP-II project has an excellent safety record. However, on May 25, 2023, a very serious construction accident occurred (involving a subcontractor for the primary civil contractor) and resulted in a shutdown of civil construction activities. The negative effects to the worker and to the project will be felt for some time to come. It is of the highest importance that projects, such as PIP-II, take safety extremely seriously and not become complacent, ensure worker training is up to date, perform timely Hazard Analysis, and focus on Work Planning & Control procedures.

Another important aspect needed to successfully execute large international partnerships is Quality Control. It starts with having an integrated Quality Assurance Plan across all Partners. PIP-II achieved this by forming the Quality Control Coordination Group which is established to bring Quality representatives from each Partner together in a forum. This group meets frequently and discusses QA expectations and ensures alignment of QC activities for like contributions. Quality Control (QC) Plans are identified and tracked via a QC Plan Tracking Tool and are developed in accordance with the PIP-II Technical Review Plan. Travelers are developed across the Project to document results of QC checks. For SRF-based production projects, quality assurance and control, data traceability and record management are essential elements as are the ability to define what constitutes acceptable performance criteria. PIP-II is in the process of establishing Acceptance Criteria Documents for all critical components. The Systems Engineering Management Plan calls out for a System Acceptance Review at the partner's institution (SAR1) and again, in most cases, at FNAL upon receipt of component (SAR2). Defining the acceptance criteria beforehand alleviates last minute decisions on what is acceptable. The acceptance criteria are tied into the Technical Requirements Specification and the required performance in the Linac.

## PROJECT STATUS

PIP-II is one year into the execution phase and has achieved several important technical milestones. There have also been challenges. The Cost Weighted Design Maturity is 92% as there are still several items awaiting Final Design Review. The most important achievement has been the successful fabrication, installation, and testing of the Front-End including the ion source, Low Energy Beam Transport (LEBT), Radio Frequency Quadrupole (RFQ), Medium Energy Beam Transport (MEBT), Half Wave Resonator (HWR) and prototype Single Spoke Resonator (SSR1) cryomodule. This configuration achieved 17 MeV energy with the beam parameters needed for LBNF/DUNE operations. The plan is to utilize all these components in the Linac. The operation of this part of the machine in the Injector Test Facility (PIP2IT) resulted in important feedback to the design teams and valuable experience for the operators.

The PIP-II scope of work remaining is still large, as there are many components needed to be fabricated and tested, including production cryomodules, magnets (both superconducting and conventional), power supplies (for cavities and magnets), instrumentation, controls, and cryogenic compo-nents. There are also the extensive civil construction activities to restart and complete. A partial list of remaining components is shown in Table 4.

Table 4: PIP-II By the Numbers

| Component | Number Required |
|---|---|
| SSR1 cryomodules | 2 |
| SSR2 cryomodules | 8 |
| LB650 cryomodules | 10 |
| HB650 cryomodules | 4 |
| 7 kW SSA | 16 |
| 20 kW SSA | 35 |
| 40 kW SSA | 36 |
| 70 kW SSA | 24 |
| ORBUMP magnets | 6 |
| Gradient magnets | 6 |
| Painting magnets | 10 |
| Gamma-t Quad magnets | 18 |
| Superconducting solenoids | 33 |
| Beam Current Monitors | 48 |
| Bean Position Monitors | 126 |
| Beam Loss Monitors | 235 |
| Beam Profile Monitors | 47 |

## SRF CAVITIES AND CRYOMODULES

The heart of the PIP-II Project is the SRF Linac. FNAL's history with the SRF technology along with its strong collaborative ties with other SRF institutions around the world, place it in a solid position to perform this work. The combination of an ongoing R&D effort and cryomodule production (experienced gained from LCLS-II and LCLS-II HE) gives confidence to the activities needed to construct PIP-II. For each cryomodule type, a prototype will be assembled and tested before launching into production. The LB650 CM will be built by CEA/Saclay while the HB 650 CM will be built by STFC-UKRI. FNAL will assemble and test the SSR CM (both SSR1 and SSR2 [5]). DAE labs will contribute many critical components for each type of cryomodule including cavities, cryomodule part kits, and Solid-State Amplifiers (SSA) for powering cavities.

The cryomodule teams are working well together, sharing information, and holding weekly meetings to discuss issues and status. Communication is the key to making these partnerships work.

Cryomodule testing of the HWR and SSR1 pCM at PIP2IT progressed well. [7] Measurements of maximum gradients and heat loads showed good performance while uncovering several issues that can be easily fixed before installation in the Linac, see Tables 5 and 6, and Figure 1. [6]

The HB 650 pCM testing is still ongoing but the cavities available at the time of string assembly were not of Linac quality. A second cooldown is planned after some in situ rework to reduce static heat load, but eventually the cavity string will be rebuilt. [3]

Table 5: HWR Gradients

| Cavity # | Nominal Required Gradient (MV/m) | Maximum Test Gradient (MV/m) | Notes |
|---|---|---|---|
| 1 | 1.39 | 10.8 | No Residual FE or MP |
| 2 | 1.89 | 10.9 | No Residual FE or MP |
| 3 | 3.91 | 6.5 | Coupler HV Bias Failure |
| 4 | 4.79 | 11.2 | No Residual FE or MP |
| 5 | 7.19 | 10.5 | No Residual FE or MP |
| 6 | 9.70 | 11.0 | No Residual FE or MP |
| 7 | 9.70 | 10.7 | No Residual FE or MP |
| 8 | 9.70 | 10.6 | No Residual FE or MP |

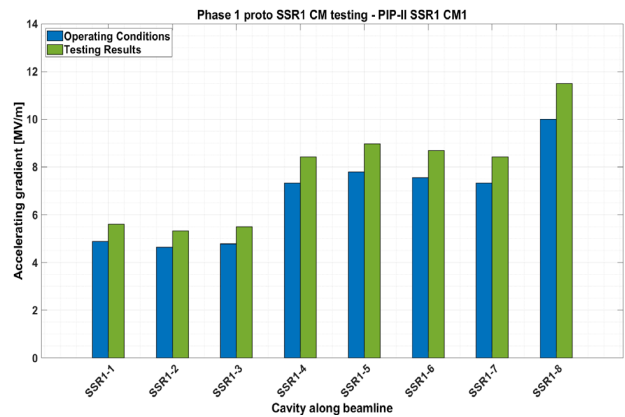

Figure 1: SSR1 pCM Gradients

Table 6: SSR1 pCM Static Heat Load

| Prototype SSR1 CM | Static Heat Loads | | |
|---|---|---|---|
|  | 2K | LTTS | HTTS |
| Design Values (W) | 25.4 | 44.2 | 150.3 |
| Measured Values (W) | 27.8±2.5 | 58.4±4.5 | 222.1±8.0 |

## CRYOGENIC SYSTEMS

The cryoplant for PIP-II is being procured from Air Liquide advanced Technologies by India/DAE. It represents the largest single In-Kind Contribution for the project. The Final Design Review was recently held at BARC, and production is underway. Other parts of the cryogenic system include the Distribution Box and Tunnel Transfer Line (both planned to be In-Kind Contributions from Poland, the Intermediate Transfer Line (FNAL procurement), and auxiliary equipment for the cryoplant (gas storage, cooling water, air systems…). The cryoplant is scheduled for delivery at the end of calendar year 2024 and should be operational in 2025.

Some key information concerning the cryo capacity
- 2.5 kW @ 2K (50% margin to baseline)
- 1.5 kW @ 5K LTTS (120% margin to baseline)
- 10.7 kW @ 40K HTTS (70% margin to baseline)

Given the cryogenic margins available, it is very important to understand any measurable level of unexpected heat loads to the system. Currently we are looking at remeasuring the SSR1 pCM heat load and lowering the HB650 pCM heat load by incorporating several small changes in thermal shorting and intercepting more heat at higher temperatures.

The cryoplant building is essentially complete and ready for the installation of various piping and electrical systems which will be done prior to installation of larger components such as compressor skids or the cold box. Power to the facility will be supplied using the main electrical feeds already in place and a refurbished transformer recovered from previously used FNAL areas. Short term use of existing equipment is a way to counteract supply chain delays,

## ACCELERATOR COMPLEX UPGRADES

Since PIP-II injects into the existing Booster which then ties to the Recycler Ring and the Main Injector, there are multiple upgrades that must be done to prepare for the new beam parameters. These upgrades include new Booster injection magnets, ORBUMP magnets and all associated power supplies. collimators, and a beam absorber. One major change is that the Booster is increasing its repetition rate from 15 Hz to 20 Hz, which means faster ramp up. A test station has been set up to test critical components at the new rep rate.

In the Recycler Ring, there is a need to do slip stacking at the 20 Hz rate. New 53 MHz cavities that can operate in CW mode are also being installed.

In the Main Injector, the upgrades include an increase in RF power by installing a second power tube in each of the Main Injector RF stations. Each of the modulators will be upgraded to accommodate another tube and the solid-state drivers in each station will be upgraded with additional power to drive the additional tube. Tubes are being supplied by Communications and Power Industries (CPI) and are arriving ahead of schedule. The first prototype modulator has been upgraded and is operational, as pictured in Figure 2.

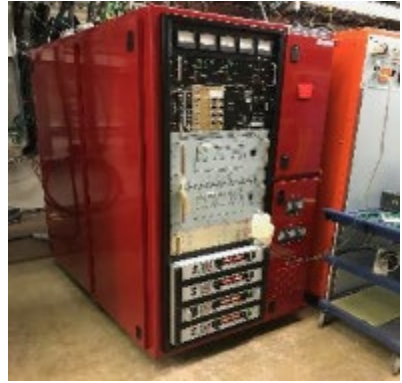

Figure 2: Main Injector Dual-Tube Modulator

## ACCELERATOR SYSTEMS

Accelerator Systems encompasses a broad range of systems including the Machine Protection System, Personnel Safety Systems, High-Power and Low-Level RF (LLRF), warm magnets and power supplies, vacuum systems, instrumentation, and controls. Essentially this work package captures all the systems outside the cold systems that are necessary to operate the Linac. The High-Power Solid-State Amplifiers (SSA) are an In-Kind Contribution from India/DAE as are some of the warm magnets. The 7kW Solid State Amplifiers (SSA) for the Spoke Resonators were used to operate the PIP2IT Front-End test. A single 40kW SSA was used to test cavities for the HB650 pCM. A total of 111 SSAs are required for the Linac. These will be procured and fabricated in India and then shipped to FNAL for installation. Acceptance criteria are being developed to ensure the components meet specification and can be NRTL certified.

The LLRF system is a collaboration amongst FNAL, LBNL, SLAC, JLab, Poland (WUT and LUT), and commercial companies. Operating the system at the test stand for the HB650 pCM has given confidence that it will be able to perform as expected. This work builds on and expands the collaboration formed for the SLAC LCLS-II Project.

PIP-II will be operated using an EPICS based control system while keeping important functionality from the legacy accelerator control system. While EPICS allows for incorporating a more modern and open-source set of controls tools, there are challenges inherent in integrating the two systems to seamlessly operate PIP-II with the downstream machines it will feed. Already the PIP2IT cryomodule test area is EPICS-based and will continue to be the prime development area prior to the beginning of commissioning.

Vacuum pumping of the beamline in the cold Linac proper is planned to use a combination Non-Evaporable Getter (NEG) and Sputter Ion Pumps as this appears to be the best technology suited for particle-free environments. Warm sections of the complex, including the Warm Front End and Beam Transfer line will be pumped with traditional distributed ion pumps. Insulating vacuum for the cryomodules will similarly use distributed pump stations. Fore-pumping is planned to be accomplished with 'dry' pumping technology. Controls interfaces are being developed (and implemented at PIP2IT) that are EPICS-based.

For instrumentation, many of the components are of a standard design. However, measuring beam profiles in the superconducting RF section of PIP-II requires technology that does not generate extraneous particles which would degrade cavity performance. Thus, a laser-based profile monitor system is being developed with thirteen pick-off points down the length of the Linac, see Figure 3. At PIP2IT, a fibre-based laser system successfully generated both transvers and longitudinal profiles, but its resolution and quality were limited by low signal to noise ratio. For PIP-II, a free-space laser will be implemented within a containment pipe under slight vacuum running the length of the Linac tunnel. Full design of such a system has inherent challenges, including stability and alignment of the laser light, personnel, and equipment safety in the presence of a free-space laser, and the challenges in generating the profile signals.

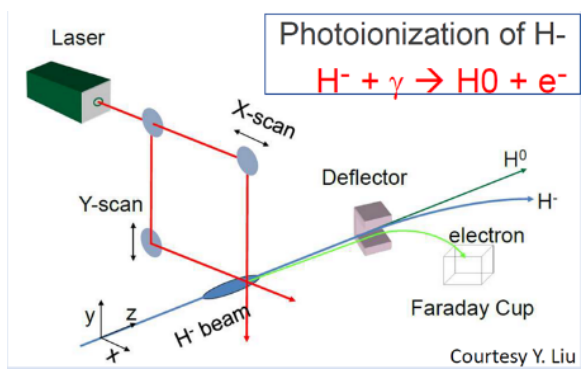

Figure 3: Laser Profile Monitor

## CONVENTIONAL FACILITIES

The Conventional Facilities section of the project focuses on civil construction and utilities. The first part of the work was started with CD-3a approval and consisted of construction of the Cryogenic Plant Building and Site Prep Work. The second part is the main Linac Complex construction, including the Front-End Building, the Linac Tunnel, and the Gallery, which houses power supplies and solid-state amplifiers. The final part is the Beam Transfer Line (BTL) construction and the connection to the existing Booster. Since the BTL section traverses the Main Ring tunnel and involves a very congested area of the complex, this is the most challenging and it has a direct impact on current beam operation.

The Site Work and Cryogenic Plant Building are essentially complete (100% and 99%, respectively), see Figure 10. The final remaining work is affected by supply chain delays with respect to electrical transformers and exceptionally long times to deliver. The Linac Complex construction was making excellent progress until a very serious construction site accident resulted in a stop work order which has paused all construction on the PIP-II site. The Booster Transfer Line is not scheduled to start construction until late Calendar Year 2025.

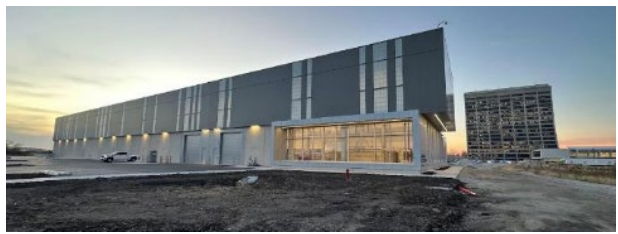

Figure 4: Cryogenic Plant Building

## CONCLUSION

The PIP-II Project moved into the execution phase with the DOE CD-3 approval in April 2022. The scheduled completion date for the early finish of the PIP-II Project is April 2029. Work is proceeding on all fronts, including SRF cryomodules, the cryogenic system, instrumentation, controls, RF amplifiers, magnets, and civil construction. Installation of Linac components is scheduled to begin in early 2025. International Partners play a key role in the machine construction and are also making good technical progress. Supply chain, and price inflation are creating challenges for the Project, but the collaboration is working to resolve them and is functioning very well together.

## ACKNOWLEDGEMENTS

Work supported by the U.S. Department of Energy, Office of Science, Office of High Energy Physics, under U.S. DOE Contract No. DE-AC02-07CH11359.